# ASSESSMENT OF SOOT FORMATION MODELS IN LIFTED ETHYLENE/AIR TURBULENT DIFFUSION FLAME

**Rohit Saini**
Department of Aerospace Engineering,
Indian Institute of Technology Kanpur,
Kanpur, Uttar Pradesh, India.
E-mail: rohits@iitk.ac.in

**Ashoke De[1]**
Department of Aerospace Engineering,
Indian Institute of Technology Kanpur,
Kanpur, Uttar Pradesh, India.
E-mail: ashoke@iitk.ac.in
Tel.: +91-512-6797863; Fax: +91-512-6797561

## ABSTRACT

*In the present study, soot formation in the turbulent lifted diffusion flame, consisting of ethylene-air is numerically investigated using three different soot modeling approaches and is comprehensively reported. For turbulence-chemistry interaction, Flamelet generated manifold (FGM) model is used. A detailed kinetics is used which is represented through POLIMI mechanism (Ranzi et al. 2012). Soot formation is modeled using two different approaches, semi-empirical two-equation approach and Quadrature methods of moments approach, where both the approaches consider various sub-processes such as nucleation, coagulation, surface growth and oxidation. The radiation heat transfer is taken into account considering four fictitious gasses in conjunction with the weighted-sum-of-gray gas (WSSGM) approach for modeling absorption coefficient. The experimental and earlier published numerical data from Köhler et al. (2012) and Blacha et al. (2011) are used for assessment of different soot modeling approaches. The discrepancies between numerical and experimental data are observed due to under-prediction of OH radicals*

---

[1] Corresponding author, Ashoke De

*concentration and poor fuel-air mixing ratios in the vicinity of the fuel jet region leading to early soot formation and the trends are unaffected after invoking radiation.*

**1. INTRODUCTION**

Urban air pollutants are becoming a potent threat and detrimental to both the environment and human health. Soot, one of them, is mainly resultant of the incomplete combustion and acts as a strong absorbent of the sunlight beside decreasing visibility and weakening of the human health. This describes, soot as a second largest contributor to the greenhouse effect subservient only to the carbon dioxide [1]. Polycyclic-aromatic-hydrocarbons (PAH) are generally formed in the fuel rich region in the downstream region of the lifted flame, eventually leading to the formation of the soot in the further downstream region. Sequentially, rich partially premixed combustion takes place in the center of the jet, followed by periphery consisting of turbulent, mixing-controlled combustion processes. Hence, the lift-off height plays an essential role in determining the spatial distribution of the maximum temperature field, therefore, also plays a primary role in the spatial variation of the soot volume fraction. Therefore, strict actions from the governmental organizations are modified for a short period to regulate the confinement of these carbonaceous particulate matters (PM). These stringent regulations consist of limiting the discharge of PM into the environment down to 4.5 mg/km for the lightweight traveler and business vehicles [2]. The challenge imposed by accurate quantification of soot formation still persists due to complex and collocated involvement of fluid mechanics, thermodynamics, radiative heat transfer and multiphase flows. Several research groups around the globe are progressing towards developing predictive models for quantification of the soot formation [3-6]. This can aide engine designers in their efforts to design the engines with better thermal efficiency by reducing the amount of combustion emissions. Nonetheless, it appears that there is still a disruption between the existing soot models and actual soot formation processes such as inception, condensation, surface growth and coagulation and others which contribute to the destruction of soot particles, such as oxidation and fragmentation. Therefore, accurate and predictive modeling of these sub-processes is required in order to develop a better understanding of soot formation in the stationary as well as dynamic reacting systems.

The bridge of numerical modeling of turbulent flames has spanned from Reynolds Averaged Navier-Stokes (RANS) computations to Large Eddy Simulation (LES) and towards Direct Numerical Simulation (DNS) as well. In a similar fashion, the advances in soot modeling have progressed from One-step soot formulation approach [7], then to Hybrid Method of Moments (HMOM) [8] and finally very expensive and detailed sectional models [9] which closely replicates the solutions of the Monte-Carlo simulations. Soot is commonly believed to be formed by coagulation of PAH species which grow by heterogeneous surface reaction with acetylene being the building block of this growth process [10]. These reactions are frequently modeled by the so-called H-Abstraction-Carbon-Addition (HACA) mechanism proposed by Frenklach and Wang [11]. Brookes and Moss [12] modeled soot volume fraction in the pilot stabilized diffusion flame burning methane/air, under which soot inception rate of particles was considered to be proportional to the local concentration of acetylene and showed good agreement with the measurements at atmospheric pressure. Primarily to include soot formation in the higher hydrocarbons, Hall et al. [13] extended the soot inception rate in the previous model, which is based on the formation of two-ringed and three-ringed aromatics from acetylene, benzene and phenyl radicals. Recently, Reddy et al. [14-16] explored empirical as well as semi-empirical soot models in Delft Flame III [17] and turbulent lifted ethylene-air flame [18] and achieved agreeable predictions as compared to experimental measurements and the published data as well. Lignell et al. [19] performed two-dimensional DNS simulations combined with semi-empirical soot modeling in non-premixed turbulent counter-flowing ethylene-air flame using acetylene as a soot precursor; later on Bisetti et al. [20] performed DNS in heptane/air turbulent counter-flow non-premixed flame using HMOM as soot model with soot inception rate, based on PAH molecules and their study has provided a brief insights in the soot particle dynamics and particle size distributions.

Sectional method approach (SM) proposed by Gelbard et al. [9] uses bins in which Number Density Function (NDF) is discretized and solved for each bin; however, accurate predictions are dependent on the number of bins required for each NDF coordinate leading to significant increase in computational cost. Another approach called, the method of the moment with interpolative closure (MOMIC) was used by Frenklach [21], which is based on the solution of the set of differential equations describing the evolution of the statistical moments ($M_n$) of the particle size distribution function (PSDF) derived from Smoluchowski's equation. It is by far the most commonly used statistics model for the prediction of the soot formation where closure to the 'n' number of moments is provided using logarithmic polynomial interpolation. Further, to incorporate multivariate distribution McGraw [22] used another interpolation approach to define the aerosol dynamics, known as Direct Quadrature Method of Moment (DQMOM). In this approach, transport equations are solved directly for the weights and locations of the delta functions. Recently, Mueller et al. [8] developed Hybrid Method of Moment (HMOM), which is a combination of MOMIC (for fractal aggregates) and DQMOM (for small spherical particulates), depicting accurate method to calculate the persistent nucleation mode. However, recently, Adhikari et al. [23] developed the transient coupling of the soot and gas phase chemistry using PSR algorithm and while comparing state of the art moment techniques i.e., HMOM and MOMIC, higher soot statistics were observed with HMOM as compared to the MOMIC. Kent & Honnery [24] performed soot volume fraction measurements in the ethylene diffusion flame with different burner configuration and Pitsch et al. [25] performed computations using only first two moments of the moment approach in conjunction with unsteady flamelet approach and obtained reasonable agreement with the experimental measurements. Furthermore, Lindstedt and Louloudi [26] computed the same flame configuration with transported PDF methodology and found that the overall effect of increased number of moments is quite modest. From previous literature, three to six moments have been found sufficient in soot modeling of premixed [6] and non-premixed flames [27].

In the present study, an attempt has been made to quantify soot predictions for ethylene/air lifted turbulent diffusion flame, utilizing semi-empirical approach and method of moments approach. The previous works carried out within the research group was primarily limited to the semi-empirical soot modeling approaches, whereas in the current work, we have assessed the results using both the approaches to estimate better predictions of soot emissions, qualitatively as well quantitatively. The prominent advantage of the method of moment approach over semi-empirical soot approach is its ability to parameterize the brief insights of soot distribution such as particle size distributions and soot particle density, later leading to the motivation of the present study. Further, only three moments have been used to predict the soot statistics to retain the computation efficiency. The broad objective of the current work is to (i) to investigate the effect of scalar quantities such as temperature, mixture fraction, and species as well as radicals formation with two different radiation approaches; (ii) to investigate the soot formation using three different soot models in combination with different OH radical determination approaches; and (iii) to investigate the effect of the soot-radiation interaction with acceptable radiation approach. The computed results are compared with previously published experimental [18] as well as the computational data [28] in order to explain the intermittencies in soot formation.

**2. THEORETICAL DESCRIPTION OF FLAMELET GENERATED MANIFOLD (FGM) APPROACH**

The FGM approach [29] the evolution of the scalar assumed in a turbulent flame are approximated by the scalar evolution in the laminar flame. In the present FGM study, one-dimensional flamelets are generated using steady laminar diffusion flamelets. Further, the two-dimensional FGM table is generated by converting scalar dissipation-mixture fraction space into a reaction progress variable-mixture fraction space. The reaction progress variable is modeled as a linear combination of the product species. The transport equation is solved for un-normalized progress variable $Y_c$ and is written as:

$$\frac{D(\overline{\rho}\tilde{Y_c})}{Dt} = \frac{\partial}{\partial x}\left(\overline{\rho D_{eff}} \frac{\partial \tilde{Y_c}}{\partial x}\right) + \overline{S_c} \qquad (1)$$

The mean source term, $\overline{S_c}$, which evaluates the turbulent flame position is modeled using the finite-rate rate obtained from the FGM library as described below:

$$\overline{S_c} = \overline{\rho} \iint P(c,Z) S_{FR}(c,Z) dc dZ \qquad (2)$$

where $S_{FR}$ is the finite-rate source term, obtained from the flamelet library, while $P$ represents the joint-PDF of mixture fraction (Z) and reaction progress variable (c). This Joint PDF '$P$' is the product of the two beta PDF's, and consists of its variances (i.e. second moments). The transport equation to model the variable of the un-normalized reaction progress variable is as follows:

$$\frac{D(\overline{\rho}\overline{Y_c''^2})}{Dt} = \frac{\partial}{\partial x}\left(\overline{\rho D_{eff}} \frac{\partial \overline{Y_c''^2}}{\partial x}\right) + C_\varphi \frac{\mu_t}{Sc_t}\left|\nabla \overline{Y_c}\right|^2 - \frac{\overline{\rho} C_\varphi}{\tau_{turb}} \overline{Y_c''^2} \qquad (3)$$

where $C_\varphi$ = 2.0 and the mean thermochemical properties are determined from the modeled PDF as:

$$\overline{\phi} = \overline{\rho} \iint P(c,Z) \phi(c,Z) dc dZ \qquad (4)$$

where $j$ denotes temperature or species fractions from the premixed flamelet library.

## 2.1 Radiation Modeling

The medium is considered as optically thick and radiation intensity is approximated by a Fourier's truncated series expansion in terms of spherical harmonics (P1 approximation) [31] given by the following equation:

$$-\nabla\left(\frac{1}{3a_l}\nabla G\right) = a_l(4\pi i_l - G_l) \quad (5)$$

In Eq. (5), the left-hand side is the gradient of radiative heat flux and the right-hand side is the source of radiative heat added to the energy equation. In the present study, non-adiabatic PDF tables in the form of three-dimensional lookup tables solving for mean mixture fraction, mean progress variable and mean enthalpy are considered in order to include the effect of heat loss or heat gain by the species fields in the burnt product zones (c = 1). Further for non-gray radiation approach, the transport equation of the incident radiation 'G' get modified and is supplied through user-defined source term which is written as:

$$\nabla \cdot (G_l \nabla G_l) - a_l G_l + 4a_l n^2 \sigma T^4 = S_{Gl} \quad (6)$$

where 'n' is the refractive index of the medium and $S_{Gl}$ is the user-defined-source term which is substituted into the energy equation to account for heat sources due to radiation. The quantity '$a_\lambda$' is the effective absorption coefficient accounting for the mixture of soot and an absorbing (radiating) gas, which is defined using Eq. (8):

$$a_l = a_{absorbing\_gas} + a_{soot}$$

(7)

The Weighted-Sum-of-Gray-Gases (WSSG) model, consisting of four fictitious gasses in the non-gray medium is used to calculate '$a_{absorbing\_gas}$' and further details regarding this approach can be found in Yadav et al. [32]. The effect of soot particles on radiative heat transfer is included using an averaged gray soot absorption coefficient, which is represented using Eq. (8):

$$a_{soot} = b_1 \rho Y_{soot}[1 - b_T(T - 2000)]$$

(8)

where $\rho$ is the soot density and $Y_{soot}$ is the mass fraction of soot. The value of $b_1 = 1232.4\ m^3/kg$ and $b_T \approx 4.8 \times 10^{-4} K^{-1}$ are ob9ained from the Taylor-Foster approximation [33] and Smith et al. [34], respectively.

**2.2 Soot Modeling**

In the present work, two semi-empirical approaches i.e. Brookes and Moss [12] and an extension of it by Hall et al. [13] are used in addition to Method of Moment (MOM) with interpolated closure for modeling soot. The results from these soot modeling approaches are compared extensively with experimental measurements and earlier published data. In previously published works [14-16], the use of above mentioned semi-empirical approach are studied extensively for modeling soot in 'Delft Flame III' and 'Lifted turbulent ethylene/air flame'. In this work as well, we have used same modeling strategy with the semi-empirical approach. Therefore, only a brief summary of those approaches are described here. The finer details of these semi-empirical approaches regarding the handling of the various source terms are not explicitly explained here and taken from [14-16]. But, the governing equations and various sub-processes of soot formation using Method of Moments are described in details.

### 2.2.1 Semi-empirical approach

Soot yield is calculated from transport equations solved for normalized soot radical nuclei concentration, $b_{nuc}^*$ and soot mass fraction $Y_{soot}$ and are described as follows:

$$\frac{\partial}{\partial t}(\rho Y_{soot}) + \nabla \cdot (\rho \vec{v} Y_{soot}) = \nabla \cdot \left(\frac{\mu_t}{\sigma_{soot}} \nabla Y_{soot}\right) + \frac{dM}{dt} \quad (9)$$

$$\frac{\partial}{\partial t}(\rho b_{nuc.}^*) + \nabla \cdot (\rho \vec{v} b_{nuc.}^*) = \nabla \cdot \left(\frac{\mu_t}{\sigma_{soot}} \nabla b_{nuc.}^*\right) + \frac{1}{N_{norm}} \frac{dN}{dt} \quad (10)$$

where $M$ and $N$ denotes soot mass concentration and soot particle number density and are used to model inception, coagulation, growth and oxidation rates. For Moss-Brookes model, the inception rate is linearly dependent on the local concentration of the acetylene as shown in Eq. (11),

$$C_2H_2 \rightarrow 2C(s) + H_2 \quad (11)$$

whereas in the extended model by Hall et al. [13], inception rate considers the formation of two and three ringed aromatics from acetylene, benzene and phenyl radicals, shown in Eq. (12, 13).

$$2C_2H_2 + C_6H_5 \rightleftharpoons C_{10}H_7 + H_2 \quad (12)$$

$$C_2H_2 + C_6H_6 + C_6H_5 \rightleftharpoons C_{14}H_{10} + H_2 + H \quad (13)$$

The oxidation due to O2 and OH are used as suggested by Neoh et al. [35] and Lee et al. [36] respectively.

## 2.2.2 Method of moment approach:

In this section, we present the governing transport equations used to get the soot mass fractions using the method of moments approach with interpolated closure [21]. In the current work, only first three moments are considered for soot size distribution and the closure is achieved by logarithmic interpolation. The concentration moment of the particle number density function is formulated for a number of moments applied and is defined as:

$$M_n = \sum_{i=1}^{\infty} m_i^n N_i \tag{14}$$

where $M_n$ and $N_i$ represents $n$th moment of soot size distribution and particle density of the size class '$i$' respectively. Accordingly, $n$ = 0 and 1 corresponds to total particle density and total mass of the particles. In order to get the soot concentrations, the following transport equation for number of moments is solved in the physical space:

$$\frac{D(\bar{\rho} M_n)}{Dt} = \nabla \cdot \left( \frac{\mu_{eff.}}{\sigma_t} \nabla M_n \right) + S_n \tag{15}$$

The source term related to soot statistics i.e. nucleation, coagulation, surface growth and oxidation source term are termed as '$S_n$' in Eq. (15) and can be written as follows:

$$\frac{dM_n}{dt} = S^{nuc.} + S^{coa.} + S^{surf.Growth+Oxid.} \tag{16}$$

The computations of these source terms are discussed in detail as follows:

**Nucleation**

Nucleation can be modeled as a coagulation process between two soot precursor species. Initially, acetylene is considered as primary soot precursor and also acts as a building block in the formation of PAH molecules [37]. However, in the present study, the source doesn't include production and consumption term due to the mass transfer rate from the gas phase to the soot, i.e. dimerization, as explained by Mueller and Pitsch [8]. The nucleation source term for the $n^{th}$ moment is calculated as:

$$S_n^{nuc.} = gC_{nuc.}\sqrt{T}[X_p]^2 \qquad (17)$$

In Eq. (18), $[X_p]$ is the molar concentration of the precursor species (acetylene in this case), $\gamma$ is the sticking coefficient [38] and $C_{nuc.}$ is the constant and is calculated using the following expression:

$$C_{nuc.} = e\sqrt{\frac{4\rho k_B}{m_c N_{C,P}}}(d_p N_A)^2 \qquad (18)$$

where $\varepsilon$ is the van der Waals enhancement factor, $N_{C,P}$ is the number of carbon atoms in a precursor molecule, $m_c$ is the mass of the carbon atom, $N_A$ is the Avogadro number, and $k_B$ is the Boltzmann constant.

**Coagulation**

In the current work, a coalescent coagulation has been assumed where particles after the collision will remain a sphere with an increased diameter. The coagulation source term for the $n = 0^{th}$ moment in the moment transport equation is calculated as follows:

$$S_n^{coa.} = -0.5 \sum_{i=1}^{\infty} \sum_{j=1}^{\infty} \beta_{i,j} N_i N_j \qquad (19)$$

As the total mass of the particle remained unaffected during coagulation process, the coagulation source term for higher moments i.e. $n \geq 2$ is calculated using Eq. (16):

$$S_n^{coa.} = 0.5 \sum_{k=1}^{n-1} \binom{n}{k} \sum_{i=1}^{\infty} \sum_{j=1}^{\infty} m_i^k m_i^{n-k} \beta_{i,j} N_i N_j \qquad (20)$$

In the Eq. (19, 20), $\beta_{i,j}$ is the collision frequency and is dependent on various coagulation regimes such as continuum, free molecular and transition (intermediate of the two). The state of the coagulation process corresponding to their regimes is defined using Knudsen number. To estimate the coagulation process regime, fixed initial population of soot particles are considered in a closed domain. The particles are initially mono-dispersed and the initial number density is $10^{18} m^{-3}$ and the initial temperature is defined as 1500 K. The particles are allowed to coagulate and the number density changed with time. The coagulation takes place in free molecular regime. Fig. 1 shows the evolution of number density with time and its comparison with earlier published results of Frenklach and Harris [39].

**Surface Growth and Oxidation**

The soot formed due to nucleation acts as an initiation of the soot formation process and is followed by the net soot production through soot surface growth pathways. Earlier, state of the art soot formation models [8, 20] considers two mechanisms for surface growth i.e. acetylene addition and PAH condensation. Simultaneously, the soot particles also lose mass due to oxidation either by $O_2$ or $OH$ species. The process of surface growth and oxidation are kinetically controlled and involves higher order of complexities, thus reduced mechanisms are used to approximate these sub-processes such that to enhance the computational efficiency. HACA mechanism is used to account soot surface growth pathways. The moment source term corresponds to the surface growth due to an addition of $C_2H_2$ species and oxidation due to $O_2$ or OH species is described as follows:

$$S = k_f [C_S^*] \sum_{k=0}^{n-1} (n,k) \Delta^{n-1} M_{k+2/3} \tag{21}$$

where S is any source for the surface growth due to $C_2H_2$ or oxidation due to $O_2$ and OH. $\Delta$ represents the number of carbons atoms added or removed. $k_f$ is the reaction rate of growth or oxidation reaction.

**Determination of OH Radical Concentration**

The two approaches namely, Equilibrium and Instantaneous approach are used in the current study to calculate OH radical concentration. In the first approach, the concentration of the OH radicals and O radicals is calculated using Eq. (22) [40,41] and Eq. (23) [42].

$$[OH] = 2.129 \times 10^2 T^{-0.57} e^{-4595/T} [O_2]^{1/2} [H_2O]^{1/2} \tag{22}$$

$$[O] = 3.97 \times 10^5 T^{-1/2} e^{-31090} [O_2]^{1/2} \tag{23}$$

In the later approach, the instantaneous value of OH radical concentration from flamelet library is used.

**2.3 Soot-turbulence interactions**

The highly non-linear interaction between soot, temperature, and species fractions is included using a single variable PDF in terms of temperature to illustrate the effect of temporal fluctuations on the net soot productions, defined as:

$$\tilde{S}_{soot} = \int PS_{soot}(T)P(T)dT \tag{24}$$

T is the instantaneous temperature and independent variables obtained from the solution of the soot transport equation are used for the construction of the PDF table. The maximum (combustion solution) and minimum (flow field solution) temperature are used as the limit for the integration. The soot-turbulence interactions are used only for the semi-empirical models.

## 3. NUMERICAL DETAILS

In the present work, soot formation in the ethylene/air turbulent lifted flame has been studied using two semi-empirical approaches along with Method of Moment (MOM) approach. The soot source terms including nucleation, coagulation, surface growth and oxidation give an important insight of soot formation processes when coupled with soot-radiation interactions. Because of the symmetry of the burner in both the cases, the present calculations are completed utilizing a two-dimensional axisymmetric configuration. The mass, momentum, energy and turbulence equations are solved using the Favre-averaged governing equations. The turbulence field is modeled using a two-equation standard $k-\varepsilon$ model with a rectification in the model constant ($C_{\varepsilon 1}$) to achieve correct jet spreading, as reported in the published literature [43]. The pressure-velocity coupling is achieved using SIMPLE algorithm and second order upwind scheme is used to discretize all the convective fluxes. All computations carried out in the present work are performed using ANSYS FLUENT-16.0 [44]. The radiative heat transfer is modeled using WSSG model and non-gray behavior is invoked through a User Defined Function (UDF) dependent on the four factious gases and weight functions are computed from Smith tables as a function of $H_2O$ and $CO_2$ partial pressures.

**4. DESCRIPTION OF TEST CASE**

The test case for the present study consists of an unconfined lifted turbulent ethylene/air flame [18] issues the fuel from a nozzle of the inner diameter of 2 mm and an outer diameter of 6 mm with a tapered sharp edged contour at the jet exit. To maintain homogeneous co-stream of oxidizer, the co-annular dry air flows through a converging section of vertical separation of 310 mm and emerges out from the exit diameter of 140 mm, encompassing the fuel jet plane. Blacha et al. [28] estimated uncertainties in scalar fields and soot measurements for this flame to be around 3% and 20 % respectively. The boundary conditions at the inlet of the burner plane are listed in Table 1. For further insights concerning the experimental setup, we might want to allude to the work by Köhler et al. [18].

**5. MODELING DETAILS AND BOUNDARY CONDITIONS**

The computational domain extends to 300D X 50D in the axial and radial directions, respectively, where D is the fuel jet diameter. The schematic of the computational domain with imposed boundary conditions is shown in Fig. 2. A grid independence study has been carried out using three non-uniform grids, namely a coarse grid with 180 (axial) X 55 (radial) cells, a medium grid with 360 (axial) X 110 (radial) cells and a fine grid with 720 (axial) X 220 (radial) cells. The exit of the computation domain is modeled as an outflow. The lateral boundary of the computational domain is used as symmetry, which enforces zero normal gradients for all scalars. The flow conditions at both the inlets are described in Table 1. The inlet velocity, turbulent kinetic energy, and turbulent dissipation are specified using a fully developed turbulent profile. The turbulence-chemistry interactions are modeled using Flamelet Generated Manifold (FGM), where the flamelets are generated using a fairly detailed mechanism containing 82 species and 1450 reactions [45]. The closure of the progress variable equation is achieved using reaction rates of the progress variable from stored computed flamelet profiles.

## 6. RESULTS AND DISCUSSION

The current section presents the predictions of soot volume fraction in the sooty turbulent diffusion lifted ethylene-air flame. Initially, grid independence study is reported, followed by a description of the structure of the flame. Later, the effect of radiative heat transfer on the soot volume fraction is investigated and intermittencies from experimental measurements are explained using source terms.

### 6.1 Grid independence study

Grid independence study is carried out using three non-uniform grids. Fig. 3 shows the centerline profiles of axial velocity and temperature with three different grids and is congruent along the axial downstream direction. All the grids follow the trend of jet spreading analogously and evolution of the temperature (b), OH mole fraction (c), and progress variable (d), with medium (with 360 (axial) X 110 (radial) cells) and finer grid (with 720 (axial) X 220 (radial) cells) are identical resulting into negligible amount of variation along the axial direction. The maximum deviation of axial velocity and the temperature is around 1.5 m/s and 85 K, respectively, between coarse and medium grid is observed. This is accompanied by the superimposed axial distribution of the OH mole fraction and the progress variable with medium and finer grids. However, the predictions from the medium grid and finer grid coincide leaving the maximum deviation to be less than 0.5% in the above-concerned parameters. Since the change between medium and fine grids are almost negligible, the medium grid i.e. 360 (axial) X 110 (radial) cells, has been chosen for the rest of the simulations.

**6.2 Structure of the flame**

First, the structure of the flame and its behavior with gray and non-gray radiation modeling approach is discussed. Later, axial and radial profiles of soot volume fraction are presented with a different approximation to compute OH radical concentration, without the inclusion of radiative heat transfer. In the end, effects of soot-radiation interactions and its impact on soot formation rates and source terms have been comprehensively investigated.

Fig. 4 depicts the axial and radial profiles of the mean temperature with the inclusion of gray and non-gray radiation. With the inclusion of gray radiation, the centerline peak temperature is reduced from ~ 2270K to ~2220K and shifted the centerline maximum temperature upstream by ~9D leading to the reduction of the flame length. In the case of non-gray radiation, the centerline maximum temperature is reduced by the difference of ~ 150K and shifted upstream by ~ 65D. A drop of ~100K is noticed in the peak centerline temperature from gray to non-gray radiation approach. This prominent difference of the peak temperature diminishes in the downstream direction as observed in the radial distribution shown in Fig. 4. However, the global maximum temperature is reduced by 70K with the inclusion of gray radiation and reduced further by ~ 274K while invoking non-gray radiation, which depicts the strong interaction of non-gray radiation with the flame environment. Thus significant improvement can be seen in Fig. 5 with the inclusion of the non-gray radiation approach.

The radial profiles of the OH species concentration are independent of the radiation approaches and distribution can be seen in the Fig. 6. The discrepancies in the magnitude of the peak OH concentration can be related to the slower kinetic rates associated with the FGM model, which may lead to the slower evolution of combustion products. The discrepancies can be attributed to the combined effects of multiple factors: (a) the radial spreading of jet at early axial location is underpredicted (Fig. 9), so the mixture fraction is not correct there, which may cause due to the turbulence model (b) the inherent assumption of 1D chemistry in FGM approach, which may come due to turbulence-chemistry interaction model, and thirdly (c) correct definition of the progress variable. Either or all of these factors may cause the observed discrepancies in OH predictions, as shown in Fig. 6; which needs detailed investigation and beyond the scope of this work. However, the trend improves while progressing in the downstream direction and location of the peak OH concentration is accurately captured at all axial distances. The experimental averaged OH mole fraction provided by Köhler et al. [18] is normalized to the maximum signal of 2500 ppm in order to compare with predictions and is shown in Fig. 7(a). The computed contour, as shown in Fig. 7(b), predicts OH mole fraction evolving at 24.3 mm and stabilized at ~ 900 K whereas experimental detection of OH starts at around 23.5 mm. The predicted lift off height is equivalent to 25 mm, which is in good agreement with the experimental value of 26 mm. The under-prediction in the gradient close to the centerline of the burner is in concurrence with the radial profiles shown in Fig. 6. During pre-combustion region shown in Fig. 7(c), the steep decrease in fuel mixture fraction is observed after 5 mm in the experimental measurements [18], whereas computed fuel mixture fraction is extended beyond 20 mm, depicting poor fuel-air mixing ratio in the proximity of the fuel jet.

The two-equation modified model is used for modeling turbulence; the model constant $C_{\varepsilon 1}$ = 1.6, is used to improve the jet spreading [43]. The effect of non-gray and gray radiation approach shown in Fig. 9, is insignificant in the axial as well as radial distribution of the axial velocity, while Figure 10(b) shows the velocity contours from current modeling approach. Jet width in the radial direction is over predicted by 20% compared to the experimental distribution shown in Figure 10(a). The currently used value of 1.6 for $C_{\varepsilon 1}$ is found to be optimum for this case and a further tuning of model constant on either side has deteriorated the results in prediction of axial jet penetration. However, Köhler et al. [18] have experienced an under-prediction of the axial velocity close to the centerline of the burner as compared to the experimental measurements which is certainly due to the input of bulk velocity of the fuel jet at the burner plate and this trend improves while progressing in the downstream direction. Similarly, slight under-prediction is noted at x = 213 mm, which is around 12 % and is acceptable considering experimental uncertainties provided by Blacha et al. [28]. Overall, the predictions are in good agreement with the experimental measurements and trend of axial velocity profiles are observed to be independent of the radiation approaches.

**6.3 Soot Predictions without radiation**

The predictions from the three different soot models i.e. Method of Moments (MOM), Moss-Brookes-Hall (MBH) and Moss-Brookes (MB), have been discussed. The dominance of gas radiation on the soot formation i.e. soot-radiation interaction has been excluded in this section. Fig. 11 exhibits the centerline peak soot volume fraction distribution predicted by the method of moments is equivalent to 0.4 ppm with OH being computed using equilibrium approach, whereas experimental [18] value is 0.54 ppm and the value predicted from sectional method [28] is 0.59 ppm. The peak value computed by both the methods i.e. method of moments as well as sectional method [28], is shifted upstream by approximately 85D due to early soot formation prediction. MB and MBH models show highly under-predicted values of soot volume fraction in the axial distribution as depicted in Fig. 11. A similar phenomenon of early soot formation is observed in semi-empirical models but peak predicted by acetylene-based inception model, i.e. MB is further shifted upstream by 106D whereas PAH based MBH model peak location prediction remains persistent with the predictions of the method of moments and sectional method. The axial distribution shows a minimal difference in the trend with equilibrium and instantaneous OH concentration approaches. Although in the radial distribution shown in Fig. 12, the span of the soot volume fraction reduces marginally with the case of instantaneous approach of OH calculation, but the maximum difference is still under 0.012 ppm. The behavior of the shift in soot volume fraction towards the fuel jet becomes more evident with two folds increase in soot volume fraction at an axial distance of 63 mm. The trend of the distribution reduces while progressing in the downstream direction further clarifies the early formation of soot particles in the flame.

The centerline profile of first two moments is presented in Fig. 13(c) with different OH concentration approaches. The peak location of the moments predicted by MOM model lies in the region of the maximum soot volume fraction present in the flame. The effect of OH concentration approaches have not shown a significant effect on the distribution of the zeroth moment i.e. total particle number density; however, the marginal shift of 5D in an axial distribution of the first moment, i.e. total mass of the particles, is observed. Fig. 13(a & b) depicts the centerline profiles of the source terms from MB and MBH models, respectively. It can be seen that the maximum soot formation rate is governed by the surface growth in both the models. Due to the inclusion of two-three ringed aromatics inception route via MBH model, the nucleation and surface growth rates increased by the order of two and shifted upstream towards the fuel jet. The difference between source terms using equilibrium approach and instantaneous approach is not significant, which is persistent with the soot volume fraction distributions in the axial as well as radial direction. None of the soot models is able to predict the location of maximum soot volume fraction, which is supposed to be associated with the early soot formation; however, the peak computed using the method of moments is in reasonable agreement with the measurements. A deviation of an order of magnitude is frequently reported for this sensitive parameter in the literature [46].

**6.4 Soot Predictions with non-gray radiation**

The centerline distribution of the soot volume fraction with different OH concentration approaches and non-gray radiation is shown in Fig. 14. The soot-radiation interactions using MOM model show the minimal effect on the peak soot volume fraction and the location of the peak value remains unchanged. The early soot formation observed previously remains intact even after the introduction of radiation. The trend of soot volume fraction distributions with MBH and MB model remains consistent with MOM model. In the present study, only acetylene is used as a primary soot precursor, which could be one of the potential reasons for under-prediction observed in the soot volume fraction. The early formation of soot particles as compared to the measurements is apparent from the accumulation of a total mass of the soot particles (N1) in the respective region. Blacha et al. [28] have also observed the similar trend of early soot formation using the sectional method and associated the phenomenon with the high number density of smaller soot particles corresponds to smaller mean particle diameter present u the upstream region as compared to the bigger soot particles in the downstream region. In order to explain these intermittencies in soot formation, the further insight of soot statistics, i.e. soot particle size distribution has been discussed and is shown in Fig. 15. It can be clearly seen that smaller particles, i.e. up to 20 nm are present in the maximum particle number density region causing early soot growth and the process continued up to 350 mm. Further downstream, soot particle sizes start reducing due to oxidation and particles with larger diameters are formed after the peak location of the soot volume fraction. Interestingly, soot particle diameters of the larger particles are found to be increased by a factor of 1.5 after invoking of non-gray radiation approach yielding strong coupling of soot-radiation effect. Since the impact of radiation heat transfer is more prominent in the downstream direction, the reduction in temperature in the region where large particles are presented has a noticeable impact on soot growth and oxidation reactions. The effect of radiation on the inception and smaller particles, which appear in lower axial locations, has marginal effect after including radiation heat transfer. Additionally, as explained in section 6.2, the poor fuel-air mixing ratio in the proximity of the jet region can be another reason for early soot formation process in the flame. The contours of the soot volume fraction from semi-empirical models and MOM model using equilibrium approach combined with non-gray radiation model is shown in Fig. 16. It can be seen that maximum soot formation in computations is restricted to the inner annular region and is concentrated toward the centerline, whereas in the experimental soot volume fraction is well developed in the outer annular regions as well, which can be possibly due to the spatially highly fluctuating soot structures in the downstream direction. Accounting the good agreement of the jet flame sensitive criteria like lift-off height, temperature and velocity predictions using RANS method to save computational efforts, potential reason in the early soot formation can be majorly linked to the inaccurate predictions of key precursor species in the soot formation and accumulation of smaller mean particle diameters in the vicinity of the nozzle. The peak magnitude of the soot volume fraction predicted by the MOM model is in fairly reasonable agreement with the measurements and shows significant improvement over two semi-empirical models while revealing more details regarding soot particles compare to other two models.

## 7. CONCLUSIONS

This study presents an assessment of soot formation using two semi-empirical based models and method of moments for turbulent diffusion lifted flame. Acetylene is used as the primary soot precursor and two different OH concentration approaches are used to determine OH radicals. The non-gray radiation approach considering four fictitious gasses is invoked to consider radiative heat losses.

Flamelet Generated Manifold (FGM) model is used to model turbulence-chemistry interaction and successfully captured the flame lift-off height. With the strong non-gray gas radiation interactions as compared to gray radiation approach, the scalar fields are observed to be in good match with the experimental measurements, however, under-prediction in OH concentration is observed in the vicinity of the jet region which improves further in the downstream direction. The early soot formation process is observed causing the location of the peak soot volume fraction to be shifted in the upstream direction. Although the global magnitude of soot volume fractions with the method of moment model is in good agreement but spatial distributions show distinct differences compared to the experimental measurements. The semi-empirical models fail to capture the magnitude as well as spatial distribution, while the effects of soot-radiation interactions are seemed to be not that prominent as the source terms from the semi-empirical models and moments from the Method of Moment model are not altered significantly after invoking soot-radiation interaction.

**ACKNOWLEDGMENT**

Financial support for this research is provided through Aeronautical Research and Development Board (ARDB), India. The computation work has been carried out on the computers provided by IITK (www.iitk.ac.in/cc). Data analysis and article preparation have been carried out using the resources available at IITK. This support is deeply appreciated.

**NOMENCLATURE**

| | |
|---|---|
| $D_{eff}$ | effective diffusivity |
| $i$ | Radiation intensity |
| $l$ | Wavelength |
| $G_l$ | Incident radiation |
| $b_{nuc}^*$ | normalized radical nuclei concentration |
| $M$ | soot mass concentration |
| $N$ | soot particle number density |
| $n$ | moment order |
| $m_{eff}$ | effective dynamic viscosity |
| $m_i$ | mass of the 'i' particle |
| $T$ | Temperature |
| $Y_i$ | species mass fraction |
| $Y_{soot}$ | soot mass fraction |
| $a_l$ | absorption coefficient |
| $\rho$ | Density |
| $\sigma_{soot}$ | turbulent Prandtl number of soot transport |
| $Z$ | mixture fraction |
| $c$ | Progress variable |
| $Y_c$ | un-normalized reaction progress variable |
| $S$ | respective source term |

**REFERENCES**

[1] Bond, T.C. and Bergstrom, R.W., Light absorption by carbonaceous particles: An investigative review, Aerosol Sci. Tech. 40(1) (2006) 27-67.

[2] Commission Regulation (EU) No 459/2012 with Amendments to Regulation (EC) No 715/2007.

[3] Appel, J., Bockhorn, H. and Frenklach, M., Kinetic modeling of soot formation with detailed chemistry and physics: laminar premixed flames of C 2 hydrocarbons, Combust. Flame 121(1) (2000) 122-136.

[4] Zhang, Q., Guo, H., Liu, F., Smallwood, G.J. and Thomson, M.J., Modeling of soot aggregate formation and size distribution in a laminar ethylene/air coflow diffusion flame with detailed PAH chemistry and an advanced sectional aerosol dynamics model, P. Combust. Inst. 32(1) (2009) 761-768.

[5] Dworkin, S.B., Zhang, Q., Thomson, M.J., Slavinskaya, N.A. and Riedel, U., Application of an enhanced PAH growth model to soot formation in a laminar coflow ethylene/air diffusion flame, Combust. Flame 158(9) (2011) 1682-1695.

[6] Mueller, M.E., Blanquart, G. and Pitsch, H., Modeling the oxidation-induced fragmentation of soot aggregates in laminar flames, P. Combust. Inst. 33(1) (2011) 667-674.

[7] Khan, I.M. and Greeves, G., A method for calculating the formation and combustion of soot in diesel engines, Heat transfer in flames 25 (1974).

**[8]** Mueller, M.E., Blanquart, G. and Pitsch, H., Hybrid method of moments for modeling soot formation and growth, Combust. Flam, 156(6) (2009) 1143-1155.

**[9]** Gelbard, F., Tambour, Y. and Seinfeld, J.H., Sectional representations for simulating aerosol dynamics, J. Colloid Interf. Sci. 76(2) (1980) 541-556.

**[10]** Richter H, Howard JB., Formation of polycyclic aromatic hydrocarbons and their growth to soot—a review of chemical reaction pathways, Prog. Energ. Combust. 26(4) (2000) 565-608.

**[11]** Frenklach, M. and Wang, H., January., Detailed modeling of soot particle nucleation and growth, In Symposium (International) on Combustion 23(1) (1991) 1559-1566.

**[12]** Brookes, S.J. and Moss, J.B., Predictions of soot and thermal radiation properties in confined turbulent jet diffusion flames, Combust. Flame 116(4) (1999) 486-503.

**[13]** Hall, R.J., Smooke, M.D. and Colket, M.B., 1997. Predictions of soot dynamics in opposed jet diffusion flames, Physical and Chemical Aspects of Combustion: A Tribute to Irvin Glassman 4 (1997) 189-230.

**[14]** Reddy, M., De, A. and Yadav, R., Effect of precursors and radiation on soot formation in turbulent diffusion flame, Fuel 148 (2015) 58-72.

**[15]** Reddy, B.M., De, A. and Yadav, R., Numerical Investigation of Soot Formation in Turbulent Diffusion Flame with Strong Turbulence–Chemistry Interaction, Journal of Thermal Science and Engineering Applications 8(1) (2016) 011001.

**[16]** Busupally, Manedhar Reddy and Ashoke De., Numerical modeling of soot formation in a turbulent C2H4/air diffusion flame, International Journal of Spray and Combustion Dynamics 8 (2016) 67-85.

[17] Qamar, N.H., Alwahabi, Z.T., Chan, Q.N., Nathan, G.J., Roekaerts, D. and King, K.D., Soot volume fraction in a piloted turbulent jet non-premixed flame of natural gas, Combust. Flame 156(7) (2009) 1339-1347.

[18] Köhler, M., Geigle, K.P., Meier, W., Crosland, B.M., Thomson, K.A. and Smallwood, G.J., Sooting turbulent jet flame: characterization and quantitative soot measurements, Appl. Phys. B 104(2) (2011) 409-425.

[19] Lignell, D.O., Chen, J.H., Smith, P.J., Lu, T. and Law, C.K., The effect of flame structure on soot formation and transport in turbulent nonpremixed flames using direct numerical simulation, Combust. Flame 151(1) 2007) 2-28.

[20] Bisetti, F., Blanquart, G., Mueller, M.E. and Pitsch, H., On the formation and early evolution of soot in turbulent nonpremixed flames, Combust. Flame 159(1) (2012) 317-335.

[21] Frenklach, M., Method of moments with interpolative closure, Chem. Eng. Sci. 57(12) (2002) 2229-2239.

[22] McGraw R., Description of aerosol dynamics by the quadrature method of moments, Aerosol Sci. Tech. 27(2) (1997) 255-265.

[23] Adhikari S, Sayre A, Chandy AJ., A Hybrid Newton/Time Integration Approach Coupled to Soot Moment Methods for Modeling Soot Formation and Growth in Perfectly-Stirred Reactors, Combust. Sci. Technol. 188(8) (2016) 1262-82.

[24] Kent, J.H. and Honnery, D., Soot and mixture fraction in turbulent diffusion flames, Combust. Sci. Tech. 54(1-6) (1987) 383-398.

[25] Pitsch, H., Riesmeier, E. and Peters, N., Unsteady flamelet modeling of soot formation in turbulent diffusion flames, Combust. Sci. Tech. 158(1) (2000) 389-406.

[26] Lindstedt, R.P. and Louloudi, S.A., Joint-scalar transported PDF modeling of soot formation and oxidation, P. Combust. Inst. 30(1) ( 2005) 775-783.

[27] Wang, H., Du, D.X., Sung, C.J. and Law, C.K., Experiments and numerical simulation on soot formation in opposed-jet ethylene diffusion flames, In Symposium (International) on Combustion 26(2) (1996) 2359-2368.

[28] Blacha, T., Di Domenico, M., Köhler, M., Gerlinger, P. and Aigner, M., Soot modeling in a turbulent unconfined C2H4/air jet flame, In 49th AIAA Aerosp. Sci. Meet. Incl. New Horizons Forum Aerosp. Expo (2011) 1-10.

[29] Oijen, J.V. and Goey, L.D., Modelling of premixed laminar flames using flamelet-generated manifolds, Combust. Sci. Tech. 161(1) (2000) 113-137.

[30] Peters, N., Laminar diffusion flamelets in non-premixed turbulent combustion. Prog, Energy Combust. Sci. 3 (1984) 319–339

[31] Howell, J.R., Menguc, M.P. and Siegel, R., Thermal radiation heat transfer, CRC press, 2010.

[32] Yadav, R., Kushari, A., Verma, A.K. and Eswaran, V., Weighted sum of gray gas modeling for nongray radiation in combusting environment using the hybrid solution methodology, Numer. Heat Tr. B-Fund. 64(2) (2013), 174-197.

[33] Taylor, P.B. and Foster, P.J., Some gray gas weighting coefficients for $CO_2$-$H_2O$-soot mixtures, Int. J. Heat and Mass Tran. 18(11) (1975) 1331-1332.

[34] Smith, T.F., Shen, Z.F. and Friedman, J.N., Evaluation of coefficients for the weighted sum of gray gases model, J. Heat Transf. 104(4) (1982) 602-608.

[35] Neoh, K. G., Howard, J.B., Sarofim, A. F., Particulate Carbon Formation during Combustion (DC Siegla and GW Smith, Eds.), Plenum Press, New York (1981) 261.

[36] Lee, K.B., Thring, M.W. and Beer, J.M., On the rate of combustion of soot in a laminar soot flame, Combust. Flame 6 (1962) 137-145.

[37] Frenklach, M., Reaction mechanism of soot formation in flames, Phys. Chem. Chem. Phys. 4(11) (2002) 2028-2037.

[38] Blanquart, G. and Pitsch, H., A joint volume-surface-hydrogen multi-variate model for soot formation, Combustion Generated Fine Carbonaceous Particles (2009) 437-463.

[39] Frenklach, M. and Harris, S.J., Aerosol dynamics modeling using the method of moments, J. Colloid Interf. Sci. 118(1) (1987) 252-261.

[40] Baulch, D.L., Cobos, C., Cox, R.A., Esser, C., Frank, P., Just, T., Kerr, J.A., Pilling, M.J., Troe, J., Walker, R.W. and Warnatz, J., Evaluated kinetic data for combustion modelling, J. Phys. Chem. Ref. Data 21(3) (1992) 411-734.

[41] Westbrook, C.K. and Dryer, F.L., Chemical kinetic modeling of hydrocarbon combustion, Prog. Energ. Combust. 10(1) (1984) 1-57.

[42] Westenberg, A.A., Kinetics of NO and CO in lean, premixed hydrocarbon-air flames, Combust. Sci. Tech. 4(1) (1971) 59-64.

[43] Pope, S. B., 1978. An explanation of the turbulent round-jet/plane-jet anomaly. AIAA Journal. 16(3), pp. 279-281.

[44] ANSYS Fluent 16.0, User's guide, <http://www.ansys.com>.

[45] Ranzi, E., Frassoldati, A., Grana, R., Cuoci, A., Faravelli, T., Kelley, A.P. and Law, C.K., Hierarchical and comparative kinetic modeling of laminar flame speeds of hydrocarbon and oxygenated fuels, Prog. Energ. Combust. 38(4) (2012) 468-501.

**[46]** Köhler M, Geigle KP, Blacha T, Gerlinger P, Meier W., Experimental characterization and numerical simulation of a sooting lifted turbulent jet diffusion flame, Combust. Flame 159(8) (2012) 2620-35.

**Table Caption List**

Table 1    The boundary conditions at the inlet of the burner as according to the experimental conditions [18].

**Figure Captions List**

Fig. 1    Evolution of Number density for free molecular regime. Line-current predictions, symbols-published results of Frenklach and Harris [39].

Fig. 2    Schematic of the computation domain used in the present study with imposed boundary conditions.

Fig. 3    Figure 3. Centerline profiles of (a) axial velocity, (b) temperature, (c) OH mass fraction, and (d) progress variable, with three different non-uniform grids used for grid-independence study.

Fig. 4    Centerline and radial distribution of the temperature at different axial locations, using gray and non-gray radiation approaches; solid symbols correspond to measurements [18].

Fig. 5    Contours of the temperature in the turbulent lifted flame with (a) no radiation, (b) gray radiation and (c) non-gray radiation approach.

Fig. 6    Radial distributions of the OH mole fraction at different axial location from the fuel nozzle exit, using gray and non-gray radiation approaches; a solid symbol corresponds to the measurements [18].

Fig. 8    Radial profiles of the ethylene mole fraction at three different axial locations in the vicinity of the fuel nozzle, using gray and non-gray radiation approaches; a solid symbols corresponds to the measurements

[18].

Fig. 9   Centerline and radial profiles of the axial velocity at two different axial locations from the fuel nozzle exit, using gray and non-gray radiation approaches; a solid symbols corresponds to the measurements [18].

Fig. 10   Contours showing computed axial velocity field (right side) and corresponding experimental measurements [18] on the left side.

Fig. 11   Centerline profiles of soot volume fraction with three different soot modeling approaches; plane lines are with equilibrium approach and lines with symbols are with instantaneous approach; solid symbols are measurement [18] and hollow symbols are with sectional method [28].

Fig. 12   Radial profiles of the soot volume fraction with three different soot modeling approaches; plane lines are with equilibrium approach and lines with symbols are with instantaneous approach; solid symbols are measurements [18] and hollow symbols are with sectional method [28].

Fig. 13   Centerline distributions of the source terms involved with three different soot modelling approaches, blue lines are with equilibrium approach and red lines are with instantaneous approach.

Fig. 14   Centerline profiles of the soot volume fraction using non-gray radiation approach with three different soot modeling approaches; plane lines are with equilibrium approach and lines with symbols are with instantaneous approach; solid symbols are measurement [18] and hollow symbols are with sectional method [28].

Fig. 15   Centerline distribution of the mean particle diameter, $d_p$ (nm) with two different OH concentration approaches and lines without symbols are using non-gray radiation approach whereas lines with symbols are without inclusion of radiation; hollow symbols are with sectional method [28].

Fig. 16    Contours of the soot volume fraction field where (a) experimental measurement [18]; (b) Moss-Brookes (MB) approach; (c) Moss-Brookes-Hall (MBH) approach; and (d) Method of Moment (MOM) approach using non-gray radiation and equilibrium approach.

|  | Fuel | Coflow |
| --- | --- | --- |
| Mean velocity (m/s) | 44 | 0.29 |
| Flow rate (g/min) | 10.4 | 320 |
| Temperature (K) | 298 | 312 |
| Reynolds Number | 10000 | - |

Table 1: The boundary conditions at the inlet of the burner as according to the experimental conditions [18].

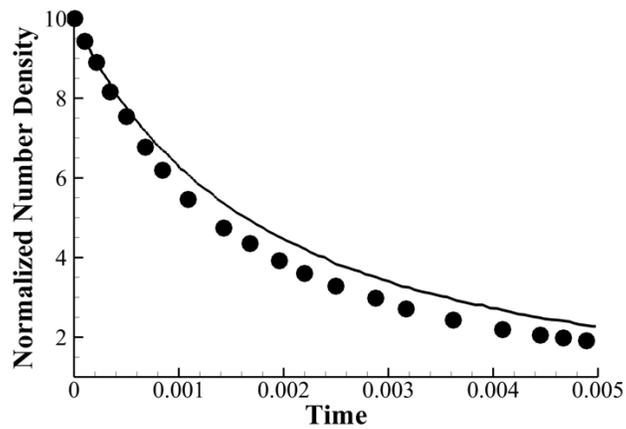

Figure 1. Evolution of Number density for free molecular regime. Line-current predictions, symbols-published results of Frenklach and Harris [39].

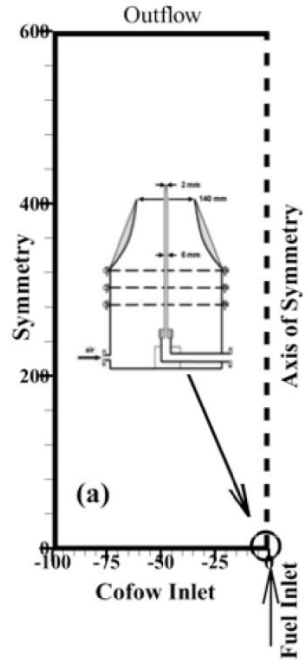

Figure 2. Schematic of the computation domain used in the present study with imposed boundary conditions.

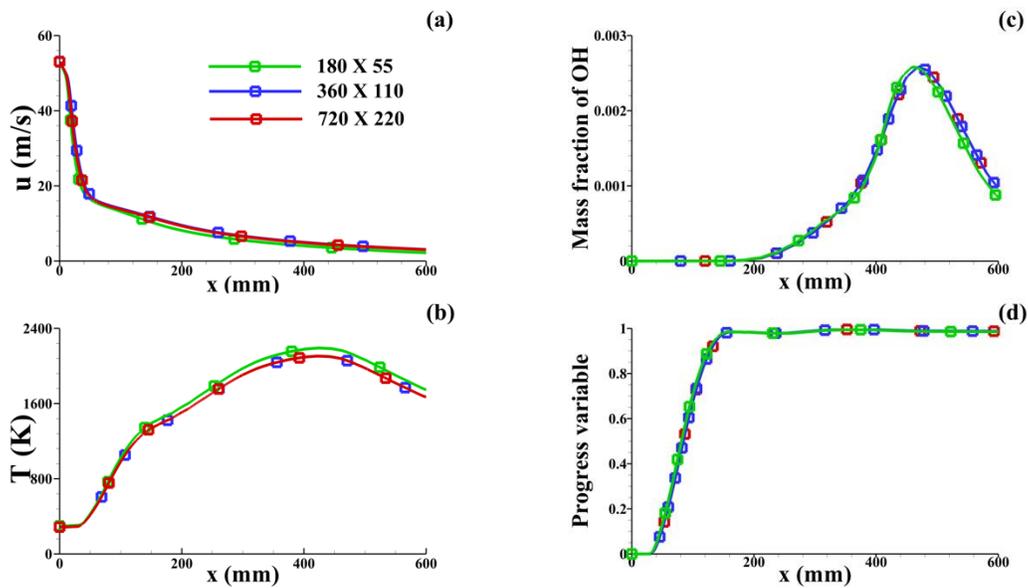

Figure 3. Centerline profiles of (a) axial velocity, (b) temperature, (c) OH mass fraction, and (d) progress variable, with three different non-uniform grids used for grid-independence study.

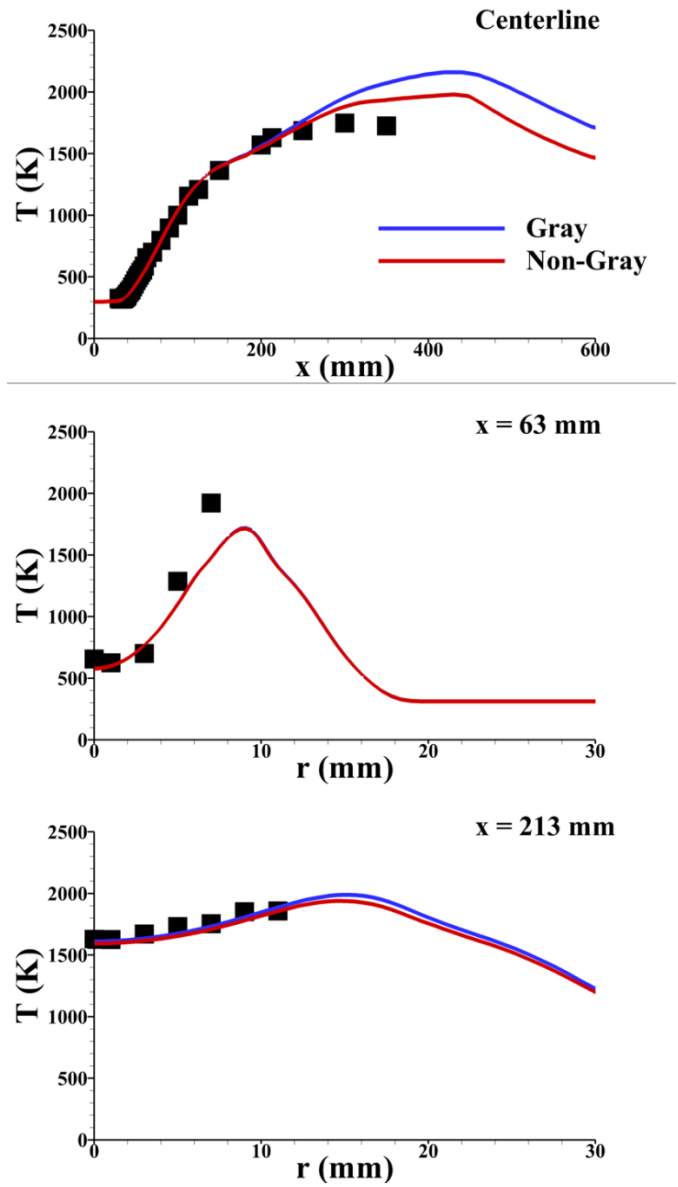

Figure 4. Centerline and radial distribution of the temperature at different axial locations, using gray and non-gray radiation approaches; symbols correspond to measurements [18].

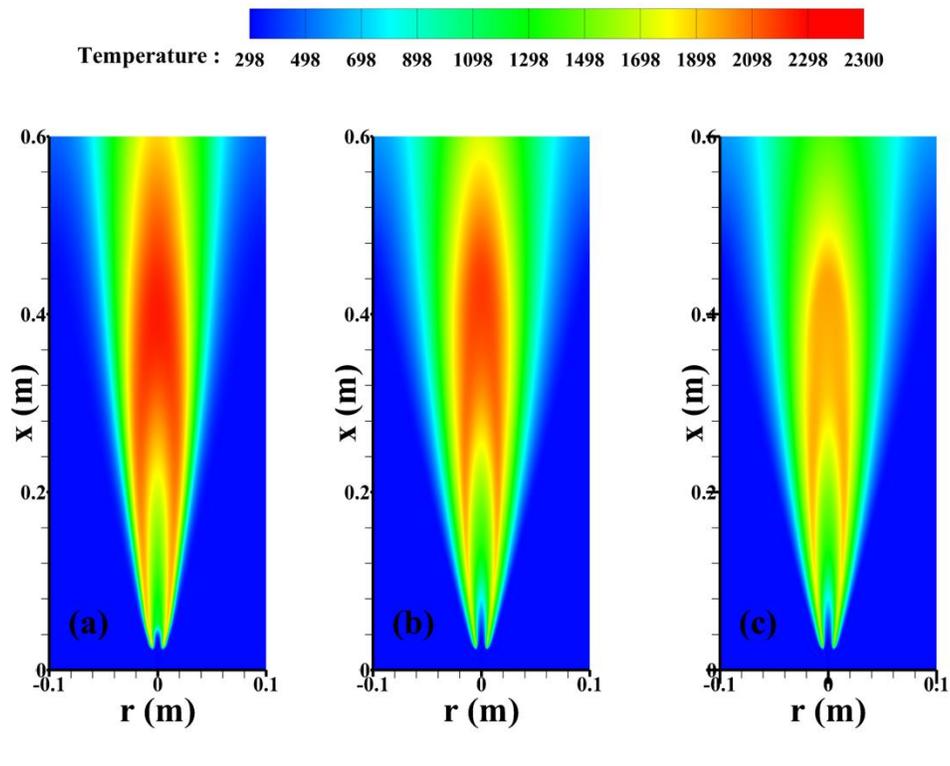

Figure 5. Contours of the temperature in the turbulent lifted flame with (a) no radiation, (b) gray radiation and (c) non-gray radiation approach.

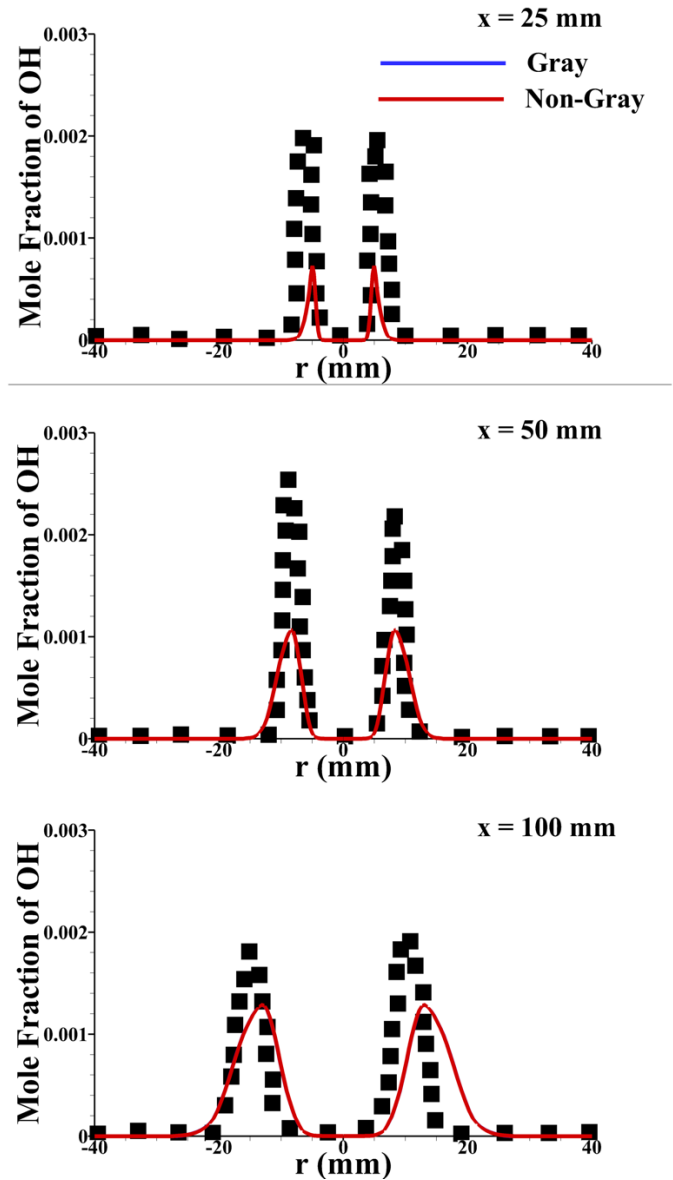

Figure 6. Radial distributions of the OH mole fraction at different axial location from the fuel nozzle exit, using gray and non-gray radiation approaches; symbols correspond to the measurements [18].

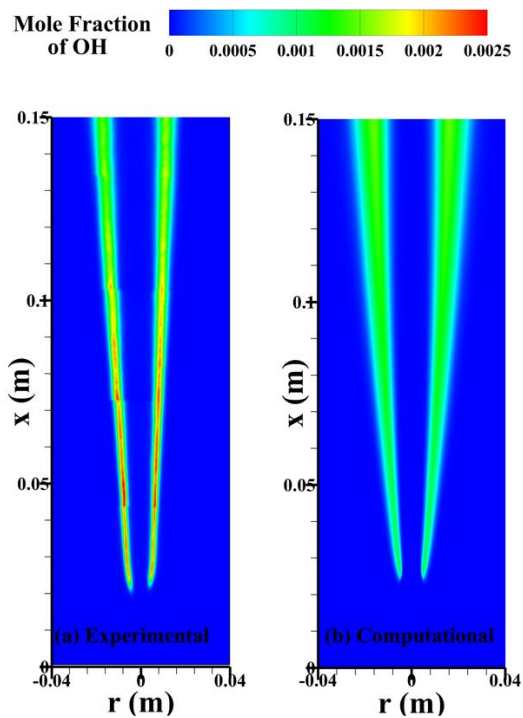 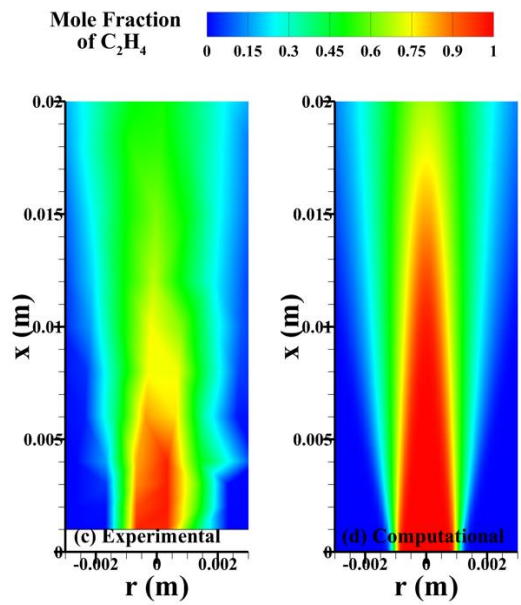

Figure 7. Experimental measurements [18] (a & c) and corresponding computed contours of the averaged mole fraction fields of (b) OH radicals and (d) $C_2H_4$ species.

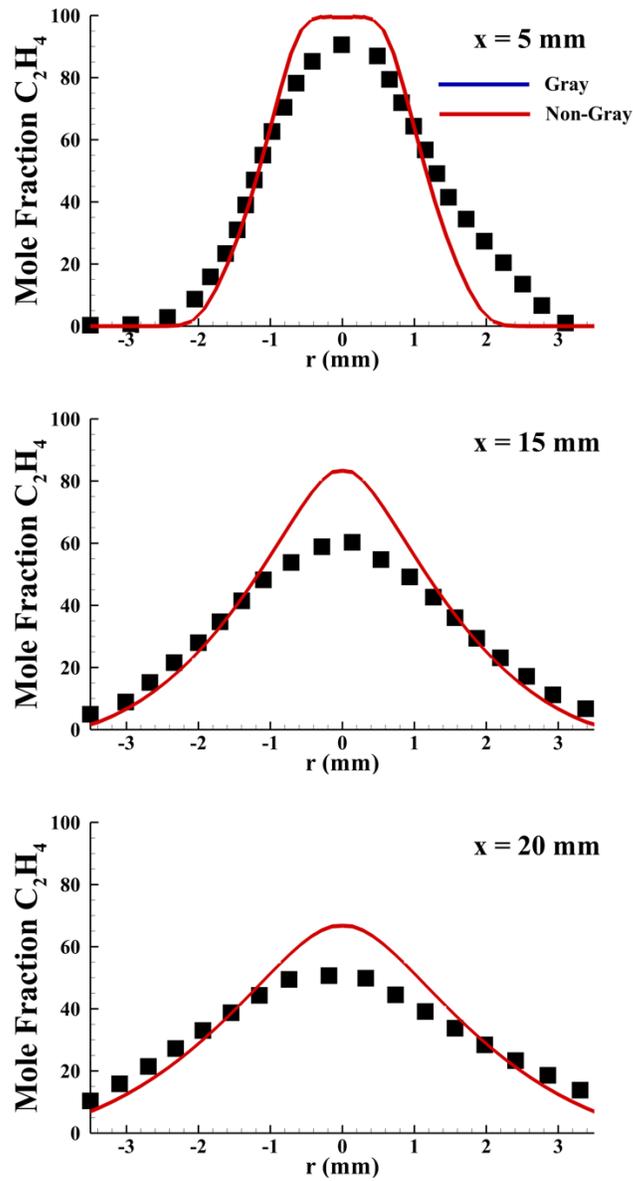

Figure 8. Radial profiles of the ethylene mole fraction at three different axial locations in the vicinity of the fuel nozzle, using gray and non-gray radiation approaches; symbols correspond to the measurements [18].

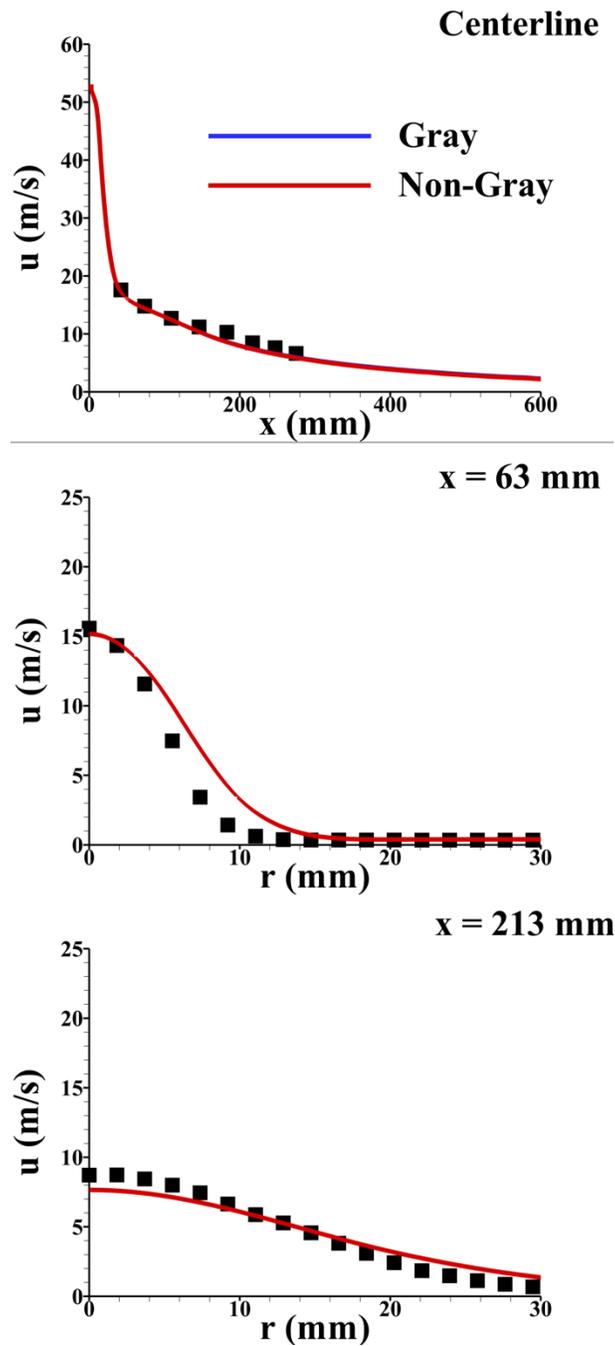

Figure 9. Centerline and radial profiles of the axial velocity at two different axial locations from the fuel nozzle exit, using gray and non-gray radiation approaches; symbols correspond to the measurements [18].

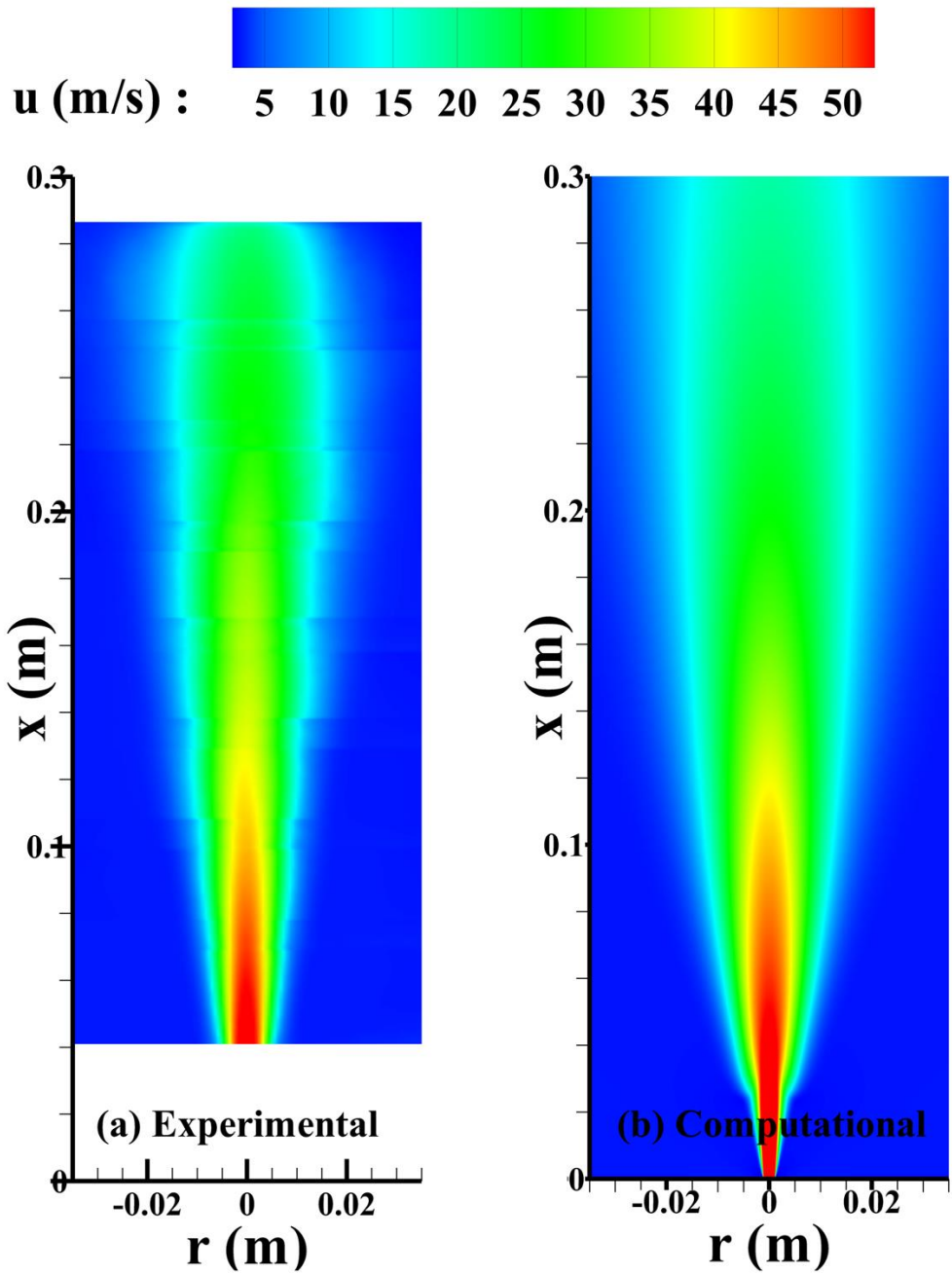

Figure 10. Contours showing computed axial velocity field (right side) and corresponding experimental measurements [18] on the left side.

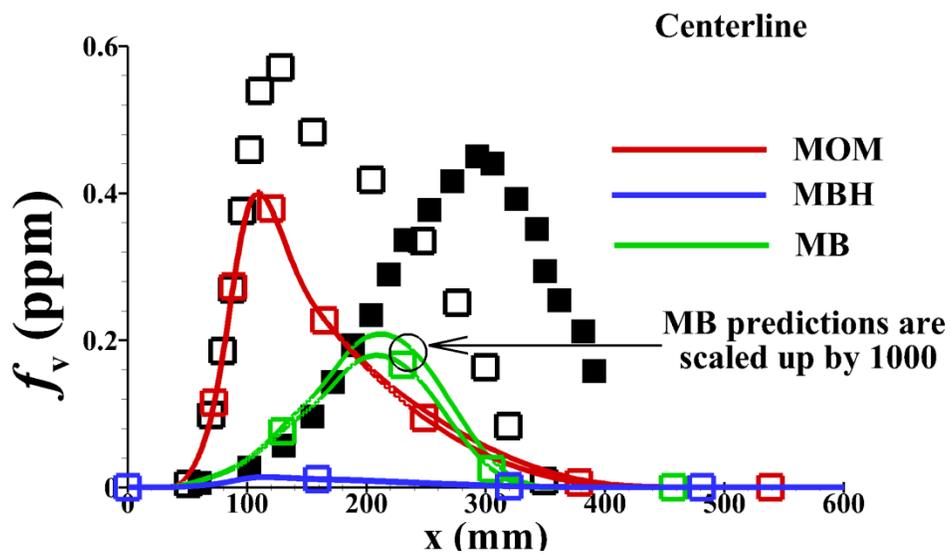

Figure 11: Centerline profiles of soot volume fraction with three different soot modeling approaches; plane lines are with equilibrium approach and lines with symbols are with instantaneous approach; solid symbols are measurements [18] and hollow symbols are with sectional method [28].

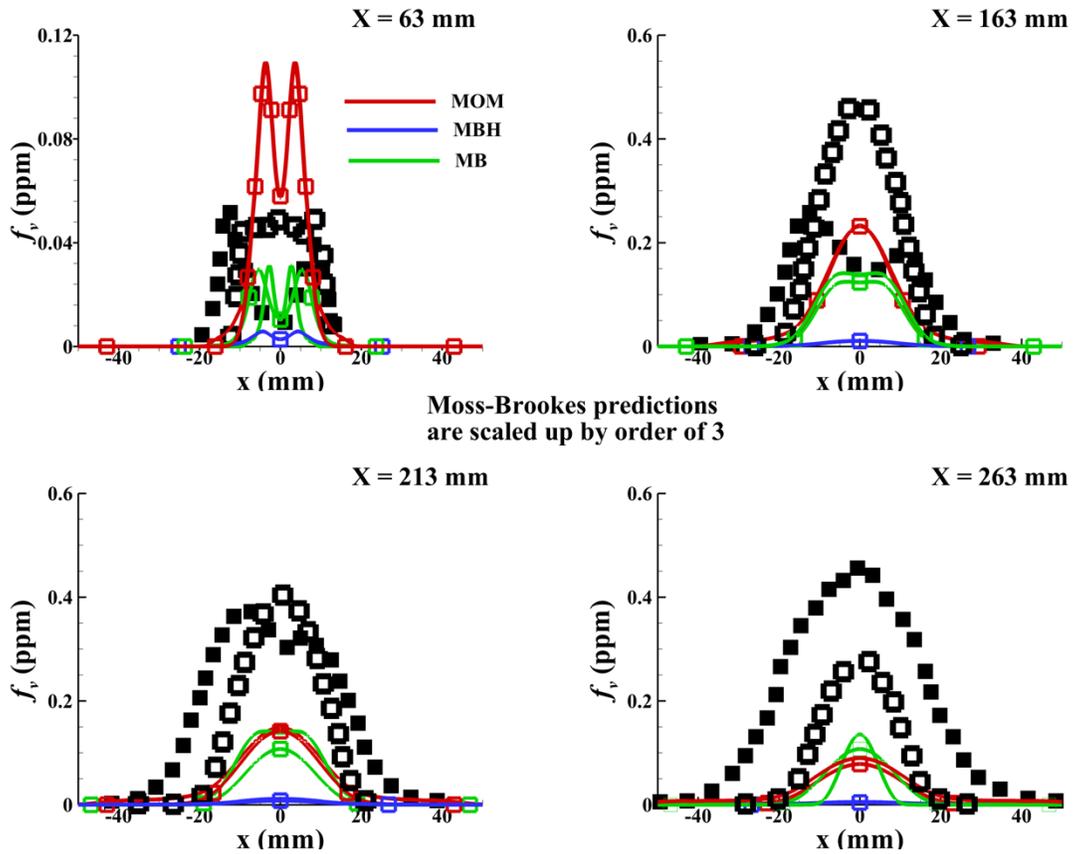

Figure 12: Radial profiles of the soot volume fraction with three different soot modeling approaches; plane lines are with equilibrium approach and lines with symbols are with instantaneous approach; solid symbols are measurements [18] and hollow symbols are with sectional method [28].

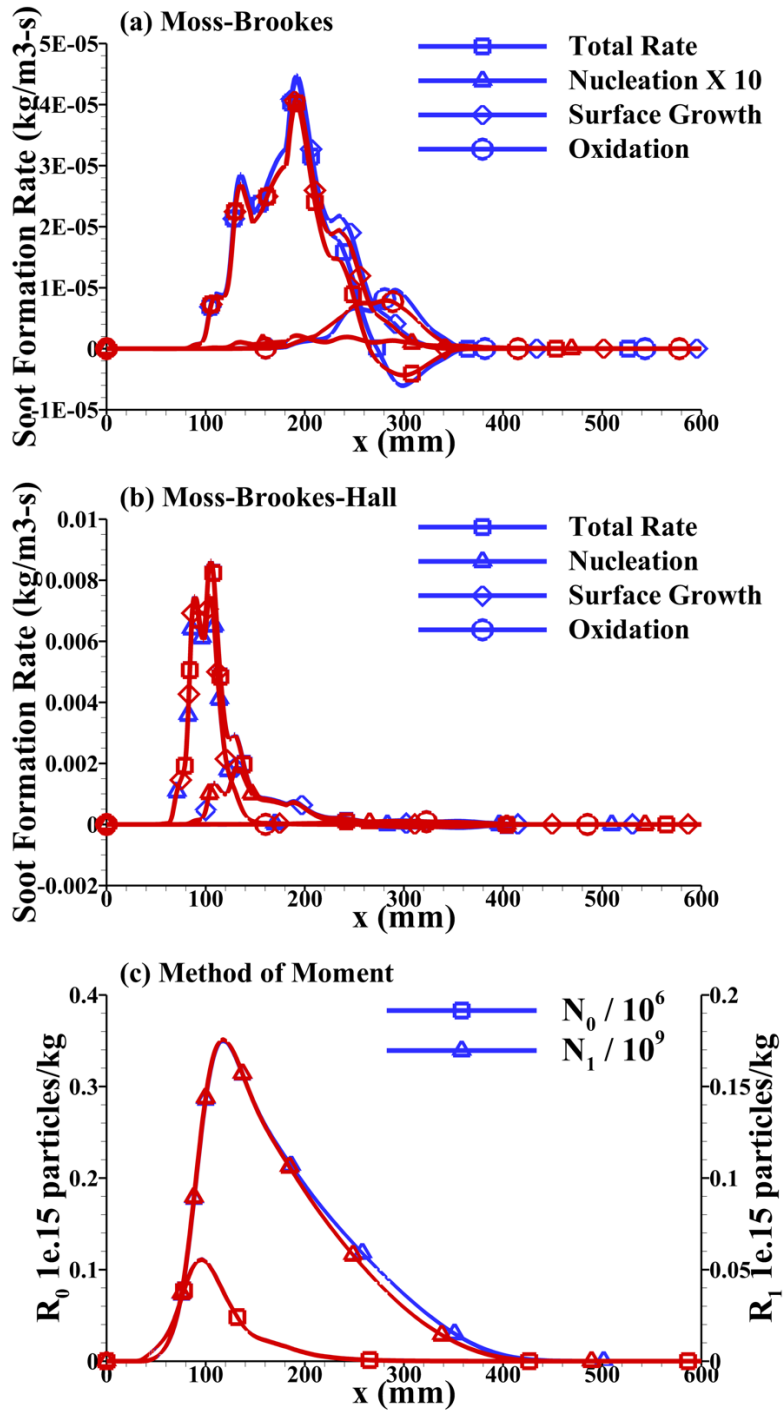

Figure 13. Centreline distributions of the source terms involved with three different soot modelling approaches, blue lines are with equilibrium approach and red lines are with instantaneous approach.

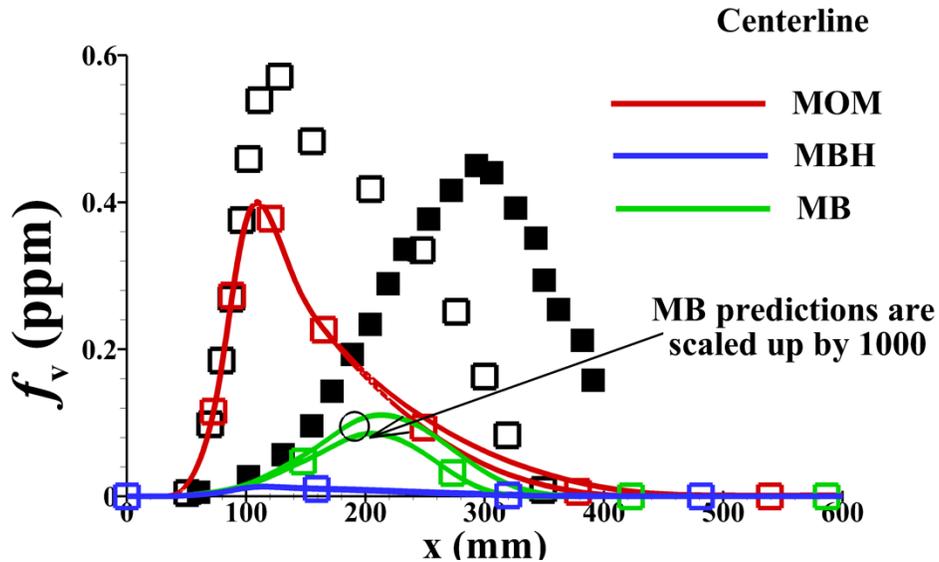

Figure 14: Centerline profiles of the soot volume fraction using non-gray radiation approach with three different soot modeling approaches; plane lines are with equilibrium approach and lines with symbols are with instantaneous approach; solid symbols are measurement [18] and hollow symbols are with sectional method [28].

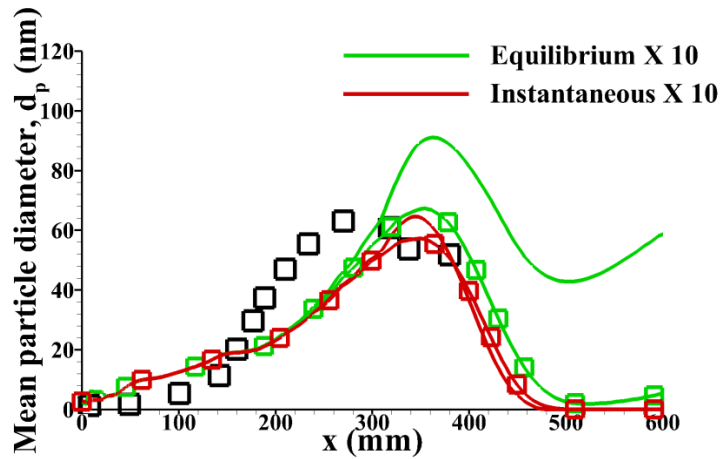

Figure 15: Centerline distribution of the mean particle diameter, $d_p$ (nm) with two different OH concentration approaches and lines without symbols are using non-gray radiation approach whereas lines with symbols are without inclusion of radiation; hollow symbols are with sectional method [28].

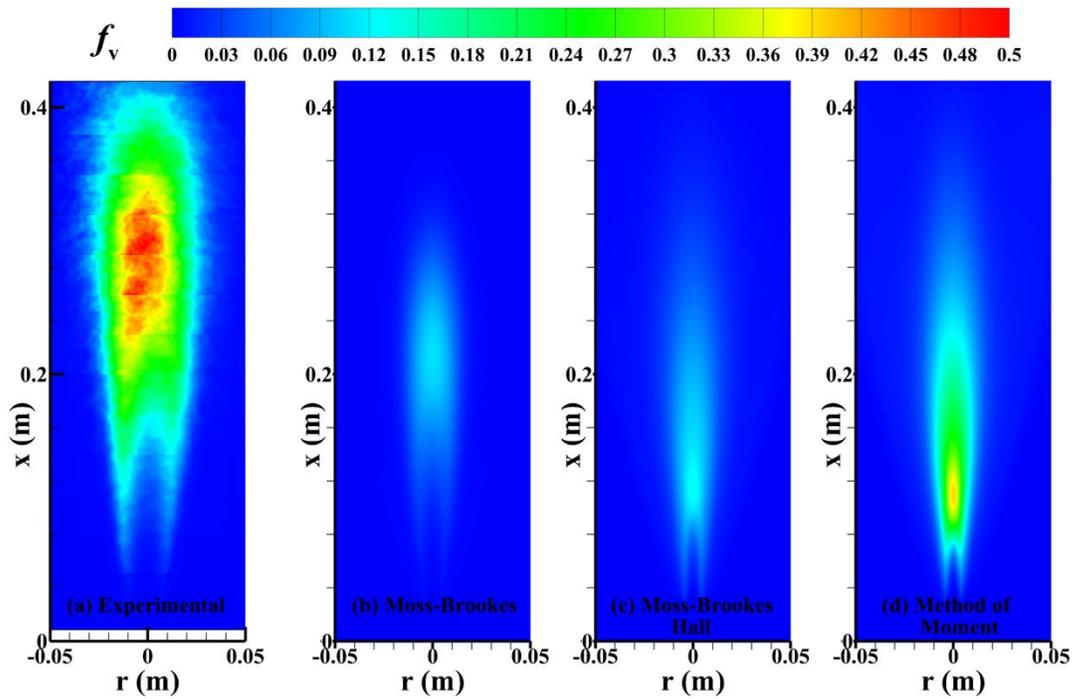

Figure 16: Contours of the soot volume fraction field where (a) experimental measurement [18];(b) Moss-Brookes (MB) approach; (c) Moss-Brookes-Hall (MBH) approach; and (d) Method of Moment (MOM) approach using non-gray radiation and equilibrium approach.